\documentclass{amsart}
\date{\today}
\def\M{\|}
\newcommand{\R}{\mathbb{R}}
\newcommand{\E}{\mathbb{E}}
\newcommand{\N}{\mathbb{N}}
\newcommand{\C}{\mathbb{C}}
\newcommand{\Z}{\mathbb{Z}}
\newcommand{\B}{\mathcal{B}}
\newcommand{\W}{\mathcal{W}}
\newcommand{\s}{\mathcal{S}}
\newcommand{\p}{\mathbb{P}}
\newcommand{\Cal}{\mathcal}

\newtheorem{Theorem}{Theorem}[section]
\newtheorem{Proposition}[Theorem]{Proposition}

\newtheorem{Lemma}[Theorem]{Lemma}

\renewcommand{\epsilon}{\varepsilon}

\begin{document}
\setcounter{section}{0}
\renewcommand{\theequation}{\arabic{section}.\arabic{equation}}
\newcounter{letters}

\title{Localization for Discrete One Dimensional Random Word Models}

\author[D.\ Damanik]{David Damanik}
\address{Department of Mathematics 253-37, California Institute of Technology,
Pasadena, CA 91125, USA; Email: damanik@its.caltech.edu}

\author[R.\ Sims]{Robert Sims}
\address{Department of Mathematics, Princeton University, Princeton,
NJ 08544, USA; Email: rsims@princeton.edu}

\author[G.\ Stolz]{G\"unter Stolz}
\address{Department of Mathematics, University of Alabama at Birmingham,
Birmingham, AL 35394-1170, USA; Email: stolz@math.uab.edu}

\thanks{D.\ D.\ was supported in part by NSF grant DMS--0227289,
G.\ S.\ by NSF grant DMS--0070343. R.\ S.\ is supported through an
NSF Postdoctoral Fellowship.}

\maketitle

\begin{abstract}
We consider Schr\"odinger operators in $\ell^2(\Z)$ whose potentials are obtained by randomly concatenating words from an underlying set $\mathcal{W}$ according to some probability measure $\nu$ on $\mathcal{W}$. Our assumptions allow us to consider models with local correlations, such as the random dimer model or, more generally, random polymer models. We prove spectral localization and, away from a finite set of exceptional energies, dynamical localization for such models. These results are obtained by employing scattering theoretic methods together with Furstenberg's theorem to verify the necessary input to perform a multiscale analysis.
\end{abstract}

%%%%%%%%%%%%%%%%%%%%%%%%%%%%%%%%%%%%%%%%%%%%%%%%%%%%%%%%%%%%%%%%%%%%%%%%%%%%%%%%%%%%%%%%%%
%                                                                                        %
%                                    Section 1                                           %
%                                                                                        %
%%%%%%%%%%%%%%%%%%%%%%%%%%%%%%%%%%%%%%%%%%%%%%%%%%%%%%%%%%%%%%%%%%%%%%%%%%%%%%%%%%%%%%%%%%

\section{Introduction}

We study one-dimensional discrete random Schr\"odinger operators
$H_\omega$, where the potential is constructed by a random
concatenation of finite words, that is, vectors in $\R^j$, $1\le
j\le m$. Our main goal is to extend known results on localization
(spectral and dynamical) for the Anderson model and the so-called
random dimer model to this more general class of random operators.
The only requirement will be that at least two words $w_1 \in
\R^{j_1}$ and $w_2 \in \R^{j_2}$ used in the construction do not
commute in the sense that $(w_1,w_2)$ and $(w_2,w_1)$ are
different vectors in $\R^{j_1+j_2}$, which avoids periodicity of
the random potential.

Generalizing known results for the Anderson model (see \cite{CKM}
and \cite{SVW} for the most general case of an arbitrary
non-trivial distribution of the coupling constant), it will be
shown that $H_\omega$ almost surely has pure point spectrum with
exponentially decaying eigenfunctions. However, as opposed to the
Anderson model, for our models the Lyapunov exponent may vanish on
a finite set of exceptional energies. That this is possible was
first observed by physicists in the example of the dimer model. It
turns out that a proof of dynamical localization requires the
exclusion of this set of energies. More precisely, we will exclude
a somewhat larger, but still finite, set of energies, at which
Furstenberg's theorem is not applicable.

In the case of the random dimer model, where an explicit analysis
of the transfer matrices allows for an exact determination of the
exceptional set, these results were proven by de Bi\`{e}vre and
Germinet \cite{Bi/Ger}. To get finiteness of the exceptional set
in our less explicit situation we employ tools from scattering
theory (reflection and transmission coefficients) and a basic fact
from inverse spectral theory (that reflection coefficients for
non-trivial scattering cannot vanish identically). These methods
were developed  in our previous work \cite{DSS} to prove
localization for continuum Anderson-type models with singular
distributions of the couplings, particularly for the Bernoulli
case. We adapt these methods to the present situation and use them
to prove positivity of the Lyapunov exponent away from the
exceptional set. One can then follow the methods used in
\cite{CKM}: After deducing H\"older continuity of the Lyapunov
exponent and the integrated density of states, one obtains a
Wegner bound and an initial length scale estimate. The latter are
the ingredients for a multiscale analysis, which yields the
localization results.

In \cite{JSS} it is shown for a subset of our models (random
polymers constructed from two words) that a generic type of the
appearing critical energies indeed leads to a breakdown of
dynamical localization. In these cases, one actually gets
superdiffusive transport. The results of \cite{Bi/Ger} and
\cite{JSS} showed that the random dimer model provides an example
of almost sure coexistence of spectral localization with dynamical
delocalization. Here we generalize \cite{Bi/Ger}, thus allowing
one to construct many other examples of this type. We note that
some of these examples will give rise to non-generic types of
exceptional energies, where the dynamical properties are not yet
known and might well be different from the behavior obtained in
\cite{JSS}.

One may also look at these models from the point of view of
``subword complexity'' of the potential---especially in the case
where the potentials take on only finitely many values. This point
of view has been discussed, for example, in \cite{d2,ds}. The
subword complexity function $p:\N \rightarrow \N$ of a given
potential $V$ (taking finitely many values) is defined as follows:
For every $n \in \N$, $p(n)$ is given by the number of distinct
subwords of length $n$ of $V$, where $V$ is regarded as an
infinite word. It is easy to see that for randomly generated
models, the complexity function is a non-random quantity, that is,
it is the same function for a full measure set of potentials. This
complexity measure is popular in many disciplines since it
discriminates nicely between periodic potentials, aperiodic
potentials with long-range order, and random potentials. For
example, a potential has a bounded complexity function $p$ if and
only if it is periodic and, on the other extremal end,
(Bernoulli-type) Anderson models have maximal word complexity
(since, almost surely, every possible word occurs). Heuristically,
a reduction of complexity should correspond to a trend from
localization to delocalization. Finite length (larger than $1$) of
the building blocks in random word models introduces some local
correlation into the potential and thus reduces complexity
compared to the Anderson model. While the trend to delocalization
is not apparent on the spectral level, it shows through the result
of \cite{JSS} on the dynamical level.

In Section~2 we define random word models, discuss some special
cases, and state our main results. Section~3 provides a formula
which relates the Lyapunov exponent of $H_\omega$ to the Lyapunov
exponent of products of independent unimodular matrices (the word
transfer matrices), thus making Furstenberg's theorem applicable
to our model. Sections~4 to 6 express the exceptional set of
energies in scattering theoretic terms, show its finiteness, and
prove positivity of the Lyapunov exponent away from the
exceptional set. Here we closely follow arguments from \cite{DSS}.
The combinatorial Lemma~\ref{lem:comb} allows us to apply a fact
from inverse spectral theory (which enters through
Lemma~\ref{zero}: Potentials can be reconstructed from
$m$-functions). In Section~7 we briefly indicate the main steps in
the remaining proof of localization, which follows the method from
\cite{CKM} with minimal changes. Due to the varying length of
words, the dynamical system underlying our random operators is a
generalization of the two-sided shift in infinite product spaces.
We include a detailed proof of the ergodicity properties of this
dynamical system in an appendix.

\vspace{.2cm}

\noindent {\bf Acknowledgements:} The authors are grateful for an
invitation to the Mittag-Leffler Institute in Fall 2002, where
this work was completed. G.\ S.\ also acknowledges the hospitality
of Universit\'e Paris 7 and financial support of CNRS (France). We
would like to thank Elliot Lieb, Eric Rains, and Shannon Starr for
useful discussions.

\section{Models and Results}

In this section we will present the models we consider and outline
the results we obtain for them. Basically, we will study discrete
Schr\"odinger operators in one dimension whose potentials are
obtained by randomly concatenating blocks from an underlying set
of words. To do so, we will fix a set of words and a probability
measure on this set. This  probability space will then lead to a
random family of discrete Schr\"odinger operators which will be
the object of study in subsequent sections.

To begin, we define the fundamental set of words from which
we will construct our operators. Fix two parameters: $m \in \N$,
the maximum word length, and $K \in (0, \infty)$, the maximum
component value of any given word. More explicitly, let
$$
\mathcal{W} := \bigcup_{j = 1}^m \mathcal{W}_j,
$$
where $\mathcal{W}_j = [-K,K]^j$. For $j=1, \ldots, m$ let $\nu_j$
be finite Borel measures on $\mathcal{W}_j$. Assume that the
$\nu_j$ are normalized such that $\sum_{j=1}^m \nu_j(
\mathcal{W}_j) =1$. We have then that $\nu$, the direct sum of the
$\nu_j$, defined by $\nu(U) = \sum_{j=1}^m \nu_j( U_j)$, for $U=
\bigcup_{j=1}^m U_j$ with $U_j \subset \mathcal{W}_j$, is a
probability measure on $\mathcal{W}$.

We must further assume a non-triviality condition on the space
$(\mathcal{W}, \nu)$; essentially, it must contain two elements
which ``do not commute'':
$$
\mbox{ (NC) } \left\{ \begin{array}{l} \mbox{ For $i=0,1$, there
exist $w_i\in \mathcal{W}_{j_i}$, both in supp$( \nu$),}
\\
\mbox{ such that the two vectors:}\\ ( w_0 (1), w_0 (2),
\ldots, w_0 (j_0), w_1 (1), w_1 (2),
\ldots, w_1 (j_1)) \ \ \mbox{and} \\ ( w_1 (1), w_1 (2), \ldots,
w_1 (j_1), w_0 (1), w_0 (2),
\ldots, w_0 (j_0) )\\
\mbox{ are distinct.}
\end{array}
\right.
$$
Here supp($\nu$) is the topological support of $\nu$ (where
$\mathcal{W}$ carries the direct sum topology of the
topologies on $\mathcal{W}_j$).

If $w \in \mathcal{W}$ belongs to $\mathcal{W}_j$, we say that $w$ has
length $j$ and write $|w| = j$. Denote the expectation of $|w|$ by
$$
\langle L \rangle = \sum_{j = 1}^m j \nu(\mathcal{W}_j).
$$
Moreover, set
$$
\Omega_0 = \mathcal{W}^\Z, \mbox{ take } \p_0 = \bigotimes_\Z \nu
$$
on the $\sigma$-algebra generated by the cylinder sets in $\Omega_0$, and
$$
\Omega = \bigcup_{j = 1}^m \Omega_j \subset \Omega_0 \times \{1,
\ldots, m \}, \mbox{ where } \Omega_j = \{ \omega \in \Omega_0 :
|\omega_0| = j \} \times \{ 1,2, \ldots, j \}.
$$
We also define a probability measure $\p$ on $\Omega$ as follows:
For any $\p_0$-measurable $A \subset \Omega_0$ such that there is
$1 \le j \le m$ with $|\omega_0| = j$ for every $\omega \in A$, we
let for $1 \le k \le j$,
\begin{equation} \label{defp}
\p \left( A \times \{ k \} \right) = \frac{\p_0(A)}{\langle L \rangle}.
\end{equation}
This determines $\p$ uniquely on the $\sigma$-algebra in $\Omega$
generated by sets of the type $A \times \{ k \}$, see the Appendix
for details. Finally, we define shifts $T_0 : \Omega_0 \rightarrow
\Omega_0$ and $T : \Omega \rightarrow \Omega$ by
$$
(T_0 \omega)_n = \omega_{n+1}
$$
and
\begin{equation} \label{deft}
T(\omega, k) = \left\{ \begin{array}{cl} (\omega,k+1) & \mbox{ if
} k < |\omega_0| \\ (T_0 \omega , 1) & \mbox{ if } k = | \omega_0
|.
\end{array} \right.
\end{equation}
It is well known that $( \Omega_0, T_0)$ is ergodic (e.g.,
\cite[p.~49]{Petersen}). It can also be shown that $( \Omega, T)$
is ergodic. More precisely, we have the following
\begin{Proposition}\label{ergprop}
Let $J:= \{ j: \nu( \mathcal{W}_j)>0 \}$. \\
{\rm (a)} If $J$ is
relatively prime, then $( \Omega, T)$ is strongly mixing.\\
{\rm (b)} If $J$ is not relatively prime, then $( \Omega, T)$ is
ergodic, but not weakly mixing.
\end{Proposition}
As this result seems to be of some interest in its own right, and
we could not find it in the literature, we include a detailed
proof in the appendix. A special case of Proposition \ref{ergprop}
(see example (v) below) was used in \cite{JSS} without proof. If
all the words have equal length, then the proof is simple and
reduces to a discrete version of the suspension procedure
described by Kirsch for continuum random operators in
\cite{Kirsch}.

We define a family of discrete Schr\"odinger operators as follows.
For $(\omega , k) \in \Omega$, we consider the operator
$$
(H_{(\omega,k)} u)(n) = u(n+1) + u(n-1) + V_{(\omega,k)}(n) u(n)
$$
in $\ell^2(\Z)$, where the potential $V_{(\omega,k)}$ results from the concatenation of
$$
\ldots, \omega_{-1}, \omega_0, \omega_1, \omega_2, \ldots
$$
such that the origin, $n = 0$, coincides with the $k$-th
position in $\omega_0$. The operator $H_{( \omega, k)}$ is
$\Z$-ergodic, that is, $V_{( \omega, k)}(n)$ is $\p$-measurable for
every $n$, and
$$
H_{T(\omega, k)} = U H_{(\omega, k)}U^{-1},
$$
where $U$ is the shift on $\ell^2( \Z)$.

We note here that the reason for having to use the probability
space $\Omega$ rather than the more trivial product space
$\Omega_0$ is the fact that words have varying length. In essence,
sequences in $\Omega$ have the additional property that the
position of their zeroth component has also been ``randomized''
relative to the origin. If all words have the same length, that
is, $\nu(\mathcal{W}_{\ell}) = 1$ for some $\ell$, then in all our
considerations we can directly work with $\Omega_0$, (always
choose $V_{\omega}(0) = \omega_0(1)$ and get $H_{T_0\omega} =
U_{\ell} H_{\omega} U_{\ell}^{-1}$, where $U_{\ell}$ is the shift
by $\ell$). In this case (NC) holds whenever $\mathcal{W}$
contains at least two words.

Let us discuss a few examples that can be studied within this framework:

\begin{itemize}
\item[(i)] \textit{Standard Anderson model}. If we set $m = 1$, we
get $\Omega = \Omega_0$, $\p = \p_0$, and the potentials are just
given by independent, identically distributed random variables.
This is the one-dimensional Anderson model whose localization
properties have been studied in many papers, most generally for
arbitrary non-trivial distribution by Carmona et al.\ \cite{CKM}
and Shubin et al.\ \cite{SVW}. \item[(ii)] \textit{Generalized
Anderson model}. One can generalize this model by taking again a
sequence of independent, identically distributed random variables
with distribution supported in $A \subset [-U,U]$ and using them
as random coupling constants of a fixed single site potential.
That is, the potential is of the form
$$
V_{\omega}(n) = \sum_{k\in\Z} q_k(\omega) f(n - k\ell),
$$
where the single site potential $f: \Z \to \R$ is supported on
$\{0, \ldots, \ell-1\}$. In our notation, this means that
$\mathcal{W} = \{ \lambda w : \lambda \in A\}$, where $w =
(f(0),\ldots, f(\ell-1)) \in \R^{\ell}$. \item[(iii)]
\textit{Discrete displacement model}. Fix integers $0< m <\ell$
and $f:\Z \to \R$ supported in $\{0,\ldots,m-1\}$. Set
$$
V_{\omega}(n) = \sum_{k\in\Z} f(n-k\ell -d_k(\omega)),
$$
with discrete i.i.d.\ random variables $d_k$ taking values in
$\{0, \ldots, \ell-m\}$. This corresponds to $\mathcal{W} = \{w_0,
\ldots, w_{\ell-m}\} \subset \R^{\ell}$ with $w_0 = (f(0), \ldots,
f(m-1), 0, \ldots, 0)$, \ldots, $w_{\ell-m} = (0, \ldots, 0, f(0),
\ldots, f(m-1))$ and $\nu(\{w_j\}) = \p(d_k=j)$. (NC) holds
whenever $f\not= 0$. \item[(iv)] \textit{Random dimer model}. A
special case of the class of examples in (ii) is given by the
random dimer model, where one sets $\ell=2$, $f=\chi_{ \{ 0,1 \}
}$. This model has been studied by de Bi\`{e}vre and Germinet in
\cite{Bi/Ger}. These authors were particularly interested in the
Bernoulli case, that is, they considered the case $\mathcal{W} =
\{ (\lambda, \lambda) , (-\lambda, -\lambda) \}$ for $\lambda >
0$. This set clearly satisfies (NC). \item[(v)] \textit{Random
polymer model}. Building on \cite{Bi/Ger}, Jitomirskaya et al.\
\cite{JSS} studied random polymer models, where one considers
random concatenations of two words. Thus, $\#(\mathcal{W}) = 2$.
Rather than studying localization properties for this model, these
authors focused on proving non-trivial lower bounds on transport
in the presence of so-called critical energies, that is, energies
at which the transfer matrices associated with the two words are
both elliptic and commute. This may very well happen even if (NC)
holds, for example at the energies $E= \pm\lambda$ in the dimer
model if $\lambda<1$.
\end{itemize}

Our goal is to prove, in the general context introduced above,
spectral localization at all energies and dynamical localization
away from a finite set of exceptional energies. This recovers
known results for the examples in (i) and (iv)
(cf.~\cite{Bi/Ger,CKM,SVW}) and, more importantly, establishes new
results for the examples in (ii), (iii) and (v). Namely, assuming
the non-triviality condition (NC), we will prove the following
pair of theorems:

\begin{Theorem}[Exponential Localization]\label{exploc}
For $\p$-almost every $(\omega, k ) \in \Omega$, the operator
$H_{(\omega,k)}$ has pure point spectrum and all eigenfunctions
decay exponentially at $\pm \infty$.
\end{Theorem}

\begin{Theorem}[Strong Dynamical Localization]\label{dynloc}
There exists a finite set $M \subset \R$ such that for
every compact interval $I \subset \R
\setminus M$, and every finitely supported $\phi \in \ell^2(\Z)$, and every $p>0$,

\begin{equation} \label{sdl}
\E \left\{ \sup_{t>0} \left\M |X|^p e^{-itH}P_I(H) \phi
\right\M \right\} < \infty,
\end{equation}
where $P_I$ is the spectral projection onto $I$.
\end{Theorem}

These theorems will be proven by adapting the scattering theoretic
approach to localization developed in \cite{DSS} to the discrete
setting. More precisely, we will show that, away from a finite set
of energies, one can apply Furstenberg's theorem to yield
positivity of the Lyapunov exponent and subsequently establish the
necessary ingredients to start a multiscale analysis. It is by now
well known that the above theorems ``follow from'' multiscale
analysis.

Let us briefly compare the results stated above with those found
in \cite{JSS}, for the models in (v). While we have to exclude
certain other types of exceptional energies as well, the critical
energies studied in \cite{JSS} are a special case of the energies
included in the set $M$. Thus the results in \cite{JSS} show that
in Theorem~\ref{dynloc} the restriction to an interval $I$ outside
$M$ is generally necessary. Also, by Theorem~\ref{exploc} every
two-word random polymer model with at least one critical energy in
the sense of \cite{JSS} provides an example of a random
Schr\"odinger operator with almost sure coexistence of exponential
localization and superdiffusive transport.

As becomes clear from the methods used in this work, in particular
Furstenberg's theorem, the number of exceptional energies
decreases if the set of words, that is, the support of the measure
$\nu$, increases. One might conjecture that for suitably rich word
spaces there are no exceptional energies. The following example
shows that non-discreteness or even connectedness of supp$\,\nu$
is not sufficient to guarantee this: Choose $\mathcal{W} =
\{\lambda w: \,\lambda \in A\}$ as in example (ii) above with $w =
(-1,0,1)$. At $E=0$ this yields the word transfer matrix (see
Section~3)
$$
M(\lambda w,0) = \left( \begin{array}{cr} 0 & -1 \\ 1 & 0
  \end{array} \right),
$$
independent of $\lambda$. This implies that the Lyapunov exponent
vanishes at $0$ for any choice of $A$.

\section{The Lyapunov Exponent}

In this section we discuss the Lyapunov exponent, which is
a quantity that measures the growth of transfer matrix norms.
This growth is also related to growth/decay properties of
generalized eigenfunctions associated with the operators
$H_{(\omega,k)}$. Due to the structure of the underlying word
space, it is of some technical advantage to define two families
of transfer matrices and two Lyapunov exponents. We shall,
however, show that these two quantities are essentially the
same.

Let us first work in the abstract setting of random products of
unimodular matrices. Given $w = (w(1), w(2), \ldots, w(j)) \in
\mathcal{W}$ and $z \in \C$, we define
$$
M(w,z) = {\mathcal T}(w(j) , z) \times \cdots \times {\mathcal T}(w(1) , z),
$$
where for $a \in \R$,
$$
{\mathcal T}(a , z) = \left( \begin{array}{cr} z - a & -1 \\ 1 & 0 \end{array} \right).
$$
For every $z \in \C$, there is a number $\gamma_0(z) \in
[0,\infty)$, called the Lyapunov exponent (of the family of random
products of matrices $M(w,z)$, $w \in \mathcal{W}$) such that we
have

\begin{equation}\label{gamma_0}
\gamma_0(z) = \lim_{N \rightarrow \infty} \frac{1}{N} \ln \|
M(\omega_N,z) \times \cdots \times M(\omega_1,z) \| \; \mbox{ for
$\p_0$-a.e.\ $\omega \in \Omega_0$}.
\end{equation}

Let us now turn to the transfer matrices associated with the
operators $H_{(\omega,k)}$. If $(\omega,k) \in \Omega$, we define
$$
M_{(\omega,k)} (n,z) = {\mathcal T}(V_{(\omega,k)}(n) , z) \times
\cdots \times {\mathcal T}(V_{(\omega,k)} (1) , z)
$$
with the ${\mathcal T}$-matrices from above. Then, for every $z
\in \C$, there is a number $\gamma(z) \in [0,\infty)$, called the
Lyapunov exponent (associated with the operator family) such that
we have

\begin{equation}\label{gamma}
\gamma(z) = \lim_{n \rightarrow \infty} \frac{1}{n} \ln \|
M_{(\omega,k)}(n,z) \| \; \mbox{ for $\p$-a.e.\ $(\omega,k) \in
\Omega$}.
\end{equation}

The main purpose of this section is to show that $\gamma$ is a fixed multiple of $\gamma_0$:

\begin{Proposition}
We have for every $z \in \C$,

\begin{equation}\label{leid}
\gamma_0(z) = \langle L \rangle \gamma(z).
\end{equation}
\end{Proposition}

\begin{proof}
Let $\tilde{\Omega}_0$ be the full measure set of those $\omega
\in \Omega_0$ such that \eqref{gamma_0} holds and also
$$
\frac{1}{k} \sum_{i = 1}^k |\omega_i| \rightarrow \langle L \rangle \; \mbox{ as } k \rightarrow \infty.
$$
For $\omega \in \tilde{\Omega}_0$, it is easily seen that
$$
\lim_{n \rightarrow \infty} \frac{1}{n} \ln \|
M_{(\omega,1)}(n,z)
\| = \frac{\gamma_0(z)}{\langle L \rangle}.
$$
Since
$$
\p \{ (\omega , 1) : \omega \in \tilde{\Omega}_0 \} = \frac{1}{\langle L \rangle} > 0,
$$
we conclude from \eqref{gamma} that \eqref{leid} holds.
\end{proof}

\section{Floquet Solutions Associated With a Periodic Potential}

In this and the following section we use various facts about algebraic
functions, in particular that they have only finitely many roots. We
refer to \cite{Knopp} for their general theory.

Let $V_{{\rm per}} : \Z \rightarrow \R$ be $p$-periodic, that is,
$V_{{\rm per}}(n+p) = V_{{\rm per}}(n)$ for every $n \in \Z$. We start
by collecting some facts from Floquet theory for the periodic operator
$H_0:= \Delta + V_{{\rm per}}$, where $(\Delta u)(n) = u(n+1) +
u(n-1)$. For any $z \in \C$, let $u_N( \cdot, z)$ and $u_D( \cdot, z)$ denote the solutions of

\begin{equation} \label{pereqn}
u(n+1) + u(n-1) + V_{{\rm per}}(n) u(n) = z u(n)
\end{equation}
with $u_N(0)=u_D(1)=1$ and $u_N(1)=u_D(0)=0$. The transfer matrix of \eqref{pereqn} from $1$ to $p$ is the matrix

\begin{equation*}
g_0(z) = \left( \begin{array}{cc} u_D(p+1,z) & u_N(p+1,z) \\
u_D(p,z) & u_N(p,z) \end{array} \right) = {\mathcal T}(V_{{\rm
per}}(p),z) \times \cdots \times {\mathcal T}(V_{{\rm per}}(1),z),
\end{equation*}
which is unimodular with polynomial entries in $z$. The eigenvalues of
$g_0(z)$ are the roots of

\begin{equation} \label{evaleqn}
\rho^2 - D(z) \rho +1 =0,
\end{equation}
that is,

\begin{equation} \label{rhoeqn}
\rho_{\pm}(z) = \frac{D(z) \pm \sqrt{D(z)^2 -4}}{2},
\end{equation}
where $D(z)= \mbox{Tr}[g_0(z)]$. As roots of \eqref{evaleqn}, the
functions $\rho_{\pm}$ are algebraic with singularities at points
with $D(z)= \pm 2$.

The spectrum of $H_0$, $\sigma(H_0)$, consists of a finite number
of bands which are given by the set of real energies $\lambda$ for
which $|D(\lambda)| \leq 2$. Let $(a,b)$ be a stability interval
of $H_0$, that is, a maximal interval such that $|D(\lambda)| < 2$
for every $\lambda \in (a,b)$. For real $\lambda \in (a,b)$, one
has that

\begin{equation*}
\rho_{\pm}(\lambda) = \tfrac{1}{2} \left( D(\lambda) \pm i
\sqrt{4- D(\lambda)^2} \right),
\end{equation*}
$|\rho_{\pm}(\lambda)|=1$, and $\rho_-(\lambda) = \overline{
\rho_+(\lambda)}$. Let

\begin{equation*}
S:= \{ z \in \C : z = \lambda + i \eta \ \ \mbox{where} \ \ a
< \lambda < b \ \ \mbox {and} \ \ \eta \in \R \}
\end{equation*}
be the vertical strip in the complex plane containing $(a,b)$. For
$z= \lambda + i \eta \in S$, one has that the following are
equivalent:
$$
\mbox{ (i) } |\rho_{\pm}(z)| =1, \; \mbox{ (ii) } \eta = 0, \; \mbox{ (iii) } D(z) \in (-2,2).
$$
The implications ${\rm (ii)} \Rightarrow {\rm (iii)}$ and ${\rm
(iii)} \Rightarrow {\rm (i)}$ are clear. To see that ${\rm (i)}
\Rightarrow {\rm (ii)}$, assume that ${\rm (i)}$ is true for some
$z = \lambda + i \eta$, where $\eta \neq 0$. As both $\rho_{\pm}$
have modulus 1, all solutions of $\eqref{pereqn}$ are bounded. By
Weyl's alternative, however, if $\eta \neq 0$, then there exists a
solution in $\ell^2$ near $+ \infty$, the Weyl solution, while all
other solutions are unbounded. This is a contradiction.

The arguments above imply that $\rho_{\pm}$ have analytic
continuations to all of $S$. These continuations remain algebraic
and the only possible singularities occur at $a$ and $b$. In
addition, as $\rho_+(z) \rho_-(z) = {\rm det}\, g_0(z) = 1$, we
have by continuity of $| \cdot|$ that exactly one of $\rho_+$ and
$\rho_-$ satisfies $| \rho(z)|<1$ in the upper half-plane. Without
loss of generality, let us denote by $\rho_+$ the eigenvalue for
which $|\rho_+ (\lambda + i \eta)|<1$ for all $\eta > 0$ and
$\lambda \in (a,b)$. This corresponds to choosing a branch of the
square root in \eqref{rhoeqn}. Then, $\rho_-$ satisfies
$|\rho_-(\lambda+i\eta)|>1$ for all $\eta >0$ and $\lambda\in
(a,b)$. Since we also have $|\rho_{\pm}( \lambda)|=1$ for $\lambda
\in (a,b)$, it follows from the Schwarz reflection principle
(apply a fractional linear transformation) that
$|\rho_+(\lambda+i\eta)|>1$ and $|\rho_-(\lambda+i\eta)|<1$ for
all $\eta <0$ and $\lambda\in (a,b)$.

For $z \in S$, let $v_{\pm}(z)$ be the eigenvectors of $g_0(z)$
corresponding to $\rho_{\pm}(z)$ with the second component
normalized to be one, that is,

\begin{equation} \label{evec}
v_{\pm}(z) = \left( \begin{array}{c} 1 \\ c_{\pm}(z) \end{array}
\right).
\end{equation}
One may easily calculate that

\begin{equation} \label{cpm}
c_{\pm}(z) = \frac{ \rho_{\pm}(z) - u_D(p+1,z)}{u_N(p+1,z)}.
\end{equation}
Here the denominator cannot vanish for $z\in S$: Suppose that
$u_N(p+1,z)=0$. Then $z$ is an eigenvalue of the finite Jacobi
matrix with diagonal $(V_{{\rm per}}(2), \ldots, V_{{\rm
per}}(p))$ and off-diagonal elements $1$. Thus $z$ and the
solutions $u_D(\cdot,z)$ and $u_N(\cdot, z)$ are real. It follows
from $1= {\rm det}\,g_0(z) = u_D(p+1,z) u_N(p,z)$ that ${\rm Tr}
\,g_0(z) = u_D(p+1,z) + u_D(p+1,z)^{-1} \ge 2$. Therefore $z$ is
either an endpoint of a stability interval or in a gap of $H_0$.

Thus, $v_{\pm}$ are analytic in $S$. In particular, as
$u_N(p+1,\cdot)^{-1}$ is a rational function with at most a pole
at $a$ and $b$, we have that $c_{\pm}$, and therefore $v_{\pm}$ as
well, are algebraic functions with, at worst, singularities at $a$
and $b$. Let $\phi_{\pm}( \cdot,z)$ be the Floquet solutions of
\eqref{pereqn}, that is, the solutions satisfying

\begin{equation} \label{fbcs}
\left( \begin{array}{c} \phi_{\pm}(1,z) \\
\phi_{\pm}(0,z) \end{array} \right) = v_{\pm}(z).
\end{equation}

We first note that $\phi_{\pm}( \cdot, \lambda+ i \eta) \in
\ell^2$ near $\pm \infty$ if $\eta >0$, and $\phi_{\pm}( \cdot,
\lambda+ i \eta) \in \ell^2$ near $\mp \infty$ if $\eta <0$. Thus
in this setting, the Floquet solutions are the Weyl solutions.
Secondly, for fixed $n$, $\phi_{\pm}(n, \cdot)$ are algebraic
functions, which are analytic in $S$ with, at most, singularities
at $a$ and $b$ arising from the singularities in the initial
conditions $v_{\pm}$. Lastly, $\{ \phi_+( \cdot,z),
\phi_-(\cdot,z) \}$ are a fundamental system of \eqref{pereqn} for
every $z \in S$, as $\rho_+(z) \neq \rho_-(z)$ on $S$.

\section{Scattering with Respect to a Periodic Background}

Let $V_{{\rm per}}$ be a $p$-periodic potential. We will insert a
local perturbation. That is, given real numbers $W_1, \ldots,
W_m$, we define a potential $V$ by
$$
V(n) = \left\{ \begin{array}{cl} V_{{\rm per}}(n) & \mbox{ if } n
\le 0 \\ W_n & \mbox{ if } 1 \le n \le m \\ V_{{\rm per}}( n - m )
& \mbox{ if } n \ge m + 1. \end{array} \right.
$$
Consider the operator $H = \Delta + V$ in $\ell^2( \Z)$. Take $z
\in S$, where $S$ is as above, and let $u_+$ be the solution of

\begin{equation} \label{perteqn}
u(n+1) + u(n-1) + V(n) u(n) =zu(n)
\end{equation}
satisfying $u_+(n) = \phi_+(n)$ for negative $n$. Since $V(n)$ and
$V_{{\rm per}}(n)$ coincide for $n \le 0$, we get that $u_+(n) =
\phi_+(n)$ for $n \le 1$. Moreover, to the right of the
perturbation, the solution $u_+$ can be written as a linear
combination of $\phi_+ ( \cdot - m + p)$ and $\phi_- (\cdot - m +
p)$, and this identity holds then for $n \ge m$ since $V(n) =
V_{{\rm per}}(n - m + p)$ for $n \ge m + 1$. In other words, there
are numbers $a(z)$ and $b(z)$ such that

\begin{equation} \label{+jost}
u_+(n,z) = \left\{ \begin{array}{cl} \phi_+(n,z) & \mbox{for } n
\le 1 \\ a(z) \phi_+(n - m + p,z) + b(z) \phi_-(n - m + p,z) & \mbox{for } n
\ge m. \end{array} \right.
\end{equation}
Since $\phi_{\pm}$ are linearly independent for $z \in S$, this
defines $a(z)$ and $b(z)$ uniquely. The numbers $t(z) = 1/a(z)$
and $r(z) = b(z)/a(z)$ are discrete analogues of the classical
transmission and reflection coefficients, at least if $V_{{\rm
per}}=0$. In particular, vanishing of $b$ is equivalent to
vanishing of the reflection coefficient. Thus $b$ and $u_+$ take
on the role of a (modified) reflection coefficient and Jost
solution relative to the periodic background $V_{{\rm per}}$,
respectively.

Since for $\lambda \in (a,b)$, we know $\phi_-(n,\lambda)=
\overline{\phi_+ (n,\lambda)}$ from \eqref{evec}, \eqref{cpm}, and
\eqref{fbcs}, by taking $u_-$ to be the solution of
\eqref{perteqn} with $u_-(n,\lambda)= \phi_-(n,\lambda)$ for $n
\le 1$, we get that for $n \ge m$,

\begin{equation} \label{-jost}
u_-(n,\lambda) = \overline{a(\lambda)} \phi_-(n - m + p,\lambda) +
\overline{b(\lambda)} \phi_+(n - m + p,\lambda).
\end{equation}
Using constancy of the non-zero Wronskian $\phi_+(n+1) \phi_-(n) -
\phi_-(n+1) \phi_+(n)$, we arrive at the familiar relation

\begin{equation} \label{abeqn}
|a(\lambda)|^2 - |b(\lambda)|^2 = 1,
\end{equation}
for $\lambda \in (a,b)$, corresponding to $|r|^2+|t|^2=1$.

\begin{Proposition}
$a( \cdot)$ and $b( \cdot)$, defined on $S$ as above, are
algebraic functions with singularities only possible
at the boundaries of stability intervals.
\end{Proposition}

\begin{proof}
Recall that $u_+$ is the solution of \eqref{perteqn} with

\begin{equation*}
\left( \begin{array}{c} u_+(1,z) \\ u_+(0,z) \end{array} \right) =
\left(  \begin{array}{c} \phi_+(1,z) \\ \phi_+(0,z) \end{array}
\right) = v_+(z).
\end{equation*}
Thus $\left( u_+(m + 1,z),u_+(m,z) \right)^t$ is
algebraic in $S$ with singularities only possible
at $a$ and $b$, that is, the boundaries of the stability interval.
As was determined before, the same is true for both $\left(
\phi_{\pm}(p + 1,z),\phi_{\pm}(p,z) \right)^t$. By the
definition of $a(z)$ and $b(z)$, we have

\begin{equation} \label{abvec}
\left( \begin{array}{c} a(z) \\ b(z)  \end{array} \right) = \left(
\begin{array}{cc} \phi_+(p + 1,z) & \phi_-(p + 1,z) \\ \phi_+(p,z) & \phi_-(p,z)
\end{array} \right)^{-1} \left( \begin{array}{c} u_+(m + 1,z) \\ u_+(m,z)
\end{array} \right),
\end{equation}
and so we are done.
\end{proof}

\begin{Lemma}\label{zero}
If $b(\lambda)=0$ for all $\lambda \in (a,b)$, then $V = V_{{\rm per}}$.
\end{Lemma}

\begin{proof}
Suppose that $b(\lambda)=0$ for all $\lambda \in (a,b)$, and hence
all $z \in S$ by analyticity. We know then that for every $\lambda
\in (a,b)$ and $\eta>0$, the Jost solution $u_+(n,\lambda+i \eta)$
is $a(\lambda + i \eta) \phi_+(n - m + p,\lambda+i\eta)$ (for all
$n \ge m$), that is, $u_+$ is exponentially decaying in this
region of the upper half-plane. Thus $u_+$ is the Weyl solution
for the perturbed equation \eqref{perteqn}. We may therefore
calculate the Weyl-Titchmarsh $m$-function, $m_V$, for
\eqref{perteqn} on the half-line $[1,\infty)$,

\begin{eqnarray*}
m_V (\lambda + i \eta) & = & \frac{u_+(1,\lambda + i \eta)}{u_+(0,\lambda + i \eta)}\\
& = & \frac{\phi_+(1,\lambda + i \eta)}{\phi_+(0,\lambda + i \eta)}\\
& = & m_{V_{{\rm per}}}(\lambda+i\eta),
\end{eqnarray*}
where the latter is the $m$-function of \eqref{pereqn} on $[1 ,
\infty)$ (which is the same as the $m$-function of \eqref{perteqn}
on $[m + 1,\infty)$). As the $m$-functions are analytic in the
entire upper half-plane, we conclude that $m_V (z)=m_{V_{{\rm
per}}} (z)$ for all $z \in \C^+$. Thus, by standard results from
inverse spectral theory (see, e.g., \cite{t}), we conclude that $V
= V_{{\rm per}}$ on $[1,\infty)$. Since they coincide on
$(-\infty,0]$ by definition, we get $V = V_{{\rm per}}$.
\end{proof}

Thus, if we restrict our attention to the case  $V \not= V_{{\rm
per}}$, we know that $\{ \lambda \in (a,b): b(\lambda)=0 \}$ is
finite.

Now we consider a gap. Take $\alpha$ such that $- \infty \leq
\alpha < a <b$ and $( \alpha, a)$ is a maximal, non-trivial gap in
the spectrum of $H_0$ (if $a = \inf{ \sigma(H_0)}$, then $\alpha =
- \infty$). We note that there is also a gap $(b, \infty)$ where
$b = \sup( \sigma(H_0))$. The analysis of this gap is identical to
the case $\alpha = - \infty$ below, excepting that we analytically
continue to the right rather than the left. Consider the following
split strip:

\begin{equation*}
S^{\prime}:= \{ z=\lambda+i\eta: \alpha < \lambda < b, \eta \in \R
\} \setminus [ \mbox{a},\mbox{b}).
\end{equation*}
For $i=1,2$, let $\rho_i(z)$ be the branches of \eqref{evaleqn}
with $|\rho_1(z)|<1$ and $|\rho_2(z)|>1$ for all $z \in
S^{\prime}$, which are well defined since $[a,b)$ is excluded. We
first note that $\rho_1 = \rho_+$ on the upper half of $S$, but
$\rho_1= \rho_-$ on the lower half of $S$. Secondly, as before, it
can be seen that $\rho_i$ are algebraic in $S^{\prime}$. They
have, at most, singularities at $\alpha$, $a$, and $b$, and
therefore they may be continued analytically across $(a,b)$. In
particular, $\rho_i$ is the analytic continuation of $\rho_j$,
where $i,j \in \{1,2\}$ with $i \neq j$.

For $z \in S^{\prime}$, choose eigenvectors $v_i(z)=(c_i(z),1)^t$
of $g_0(z)$ corresponding to $\rho_i(z)$, analogously to
(\ref{evec}). Taking $\phi_i$ to be the solutions of
\eqref{pereqn} with $(\phi_i(1,z), \phi_i(0,z) )^t = v_i(z)$ for
$z \in S^{\prime}$, we see again that $\phi_1( \cdot, z)$ (resp.,
$\phi_2( \cdot,z)$) is in $\ell^2$ near $+ \infty$ (resp., $-
\infty$), that is, they are the Weyl solutions. Set $u_i$ to be
the Jost solutions of \eqref{perteqn} satisfying

\begin{equation}\label{jostsols}
u_i( n, z) = \left\{ \begin{array}{cc} \phi_i(n , z) & n \le 1 \\ a_i(z) \phi_i(n - m + p,z) + b_i(z) \phi_j(n - m + p,z) & n \ge m \end{array}
\right.
\end{equation}
for $z \in S^{\prime}$ and the same $i,j$ convention used above. As in \eqref{abvec} above, one sees that $a_1(z)$ and $b_1(z)$ are algebraic in $S^{\prime}$. In fact, in the upper half $S_+$ of $S$, they coincide with $a(z)$ and $b(z)$, since for $z \in S_+$, we have that $v_+(z)$ and $v_1(z)$ coincide. Thus $a_1(z)$ and $b_1(z)$ are analytic continuations of the restrictions of  $a(z)$ and $b(z)$ to $S_+$. Similarly, it is seen that $a_2(z)$ and $b_2(z)$ are analytic continuations of the restrictions of  $a(z)$ and $b(z)$ to the lower half $S_-$ of $S$.
Thus, $a_1(z)$, $b_1(z)$, $a_2(z)$, and $b_2(z)$ are algebraic and have, at most, singularities at $\alpha$, $a$, and $b$. In particular, when $V \not= V_{{\rm per}}$, they cannot vanish
identically, and hence the set

\begin{equation} \label{abroots}
\{ \lambda \in (\alpha,b):
a_1(\lambda)b_1(\lambda)a_2(\lambda)b_2(\lambda)=0 \}
\end{equation}
is finite, even in the case $\alpha = - \infty$.

\section{Furstenberg's Theorem and Positivity of the Lyapunov Exponent}

To prove positivity of $\gamma_0(\lambda)$ for $\lambda \in \R$, away from a finite set, we will
investigate properties of the transfer matrices. We assumed in (NC) that there are two words $w_0$, $w_1$ in supp($\nu$) which do not commute, and we will therefore work with the following pair of transfer matrices: the ``free'' matrix
$$
g_0(\lambda) = M(w_0 , \lambda),
$$
corresponding to the periodic problem \eqref{pereqn} and the matrix
$$
g_1(\lambda) = M(w_1 , \lambda)
$$
describing the local perturbation, corresponding to the perturbed difference equation
\eqref{perteqn}. In correspondence with the previous sections, we take $H_0$ to be the operator with $p$-periodic potential $V_{{\rm per}}$, where $p = |w_0|$, such that
$V_{{\rm per}} (n) = w_0(n)$, $1 \le n \le p$, and $H$ to be the operator generated by
the perturbed potential $V$ results from $V_{{\rm per}}$ by inserting the $m$ numbers $W_1 , \ldots , W_m$, where $m = |w_1|$ and $W_n = w_1 (n)$, $1 \le n \le m$. Of course, this convention is arbitrary and the roles of the two transfer matrices could be interchanged.

\begin{Lemma} \label{lem:comb}
Assume that {\rm (NC)} holds with non-commuting words $w_0,w_1$. Then, we have for the potentials $V_{{\rm per}}, V$ defined above,

\begin{equation}\label{potinequ}
V_{{\rm per}} \not= V.
\end{equation}
\end{Lemma}

\begin{proof}
Assume that \eqref{potinequ} fails. Then, we get that $w_0$ is a
power, that is, there is some $v$, which itself is not a power,
and some $s \ge 2$ in $\N$ such that $w_0 = v^s = v v \ldots v$.
We have then that

\begin{equation}\label{words2}
w_1 v v v  \ldots = v v v v \ldots
\end{equation}
by considering the restrictions of $V_{{\rm per}}$ and $V$ to $\N$
as one-sided infinte words. If the length of $w_1$ is not an
integer multiple of the length of $v$, we can again argue that $v$
must be a power, and hence get a contradiction. Thus, the length
of $w_1$ is an integer multiple of the length of $v$. By
inspection of \eqref{words2}, this implies that $w_1$, too, is a
power of $v$. Thus $w_1 = v^{\hat{s}}$ and hence we get a
contradiction to (NC) since $w_0 w_ 1 = v^{s + \hat{s}} = w_1
w_0$.
\end{proof}

Set $G(\lambda)$ to be the closed subgroup of ${\rm SL}(2, \R)$
generated by $\{ M(w , \lambda) : w \in \mbox{supp}(\nu) \}$. Let
$P( \R^2)$ be the projective space, that is, the set of the
directions in $\R^2$ and $\overline{v}$ be the direction of $v \in
\R^2 \setminus \{ 0 \}$. Note that ${\rm SL}(2, \R)$ acts on $P(
\R^2)$ by $g \overline{v} = \overline{gv}$. We say that $G \subset
{\rm SL}(2 , \R)$ is strongly irreducible if and only if there is
no finite $G$-invariant set in $P( \R^2)$.

It follows from Furstenberg's theorem \cite{Bougerol/Lacroix} that
$\gamma_0(\lambda) > 0$ if $G( \lambda)$ is non-compact and
strongly irreducible.

The main result of this section is the following:

\begin{Theorem}\label{gamma+}
Assume that {\rm (NC)} holds. Then there exists a finite set $M
\subset \R$, such that $G( \lambda)$ is non-compact and strongly
irreducible for all $\lambda \in \R \setminus M$. In particular,
$\gamma_0 (\lambda)>0$ for all $\lambda \in \R \setminus M$.
\end{Theorem}

\begin{proof}
The proof of Theorem~\ref{gamma+} is analogous to the proof of
Theorem~2.3 in \cite{DSS} so we only sketch the argument briefly.

When non-compact, it is known that the group $G$ is strongly
irreducible if and only if for each $\overline{v} \in P(\R^2)$,

\begin{equation} \label{3dir}
\#\{g \overline{v}: g \in G\} \geq 3;
\end{equation}
see \cite{Bougerol/Lacroix}. Note that both non-compactness
of$G(\lambda)$ and \eqref{3dir} are properties which are preserved
if the set of underlying words is enlarged. Thus one can assume
without loss of generality that $G(\lambda)$ is generated by $g_0(
\lambda)$ and $g_1(\lambda)$.

By an analysis which is almost identical to the one in \cite{DSS},
non-compactness and (\ref{3dir}) can be shown away from the roots
of $b$ (finitely many in each of the finitely many stability
intervals), the roots of $D$ (one per stability interval), the
endpoints of stability intervals, and the set in (\ref{abroots});
and hence away from a finite set.

For $\lambda$ in a gap, non-compactness follows since
$g_0(\lambda)$ has an eigenvalue of modulus larger than $1$. The
set (\ref{abroots}) is avoided to verify (\ref{3dir}).

If $\lambda$ is in a stability interval, then the crucial link to
scattering coefficients is given by the fact that $G(\lambda)$ is
conjugate to the group $\tilde{G}(\lambda)$ generated by the two
matrices
$$
\tilde{g}_0(\lambda) = \left( \begin{array}{rc} \cos \omega &
\sin \omega \\ - \sin \omega & \cos \omega \end{array}
\right), \;
s(\lambda) = \left( \begin{array}{rr}
{\rm Re}[a(\lambda)+b(\lambda)] &
{\rm Im}[a(\lambda)+b(\lambda)]
\\ -{\rm Im}[a(\lambda)-b(\lambda)]
& {\rm Re}[a(\lambda)-b(\lambda)] \end{array} \right),
$$
where $\omega = \omega(\lambda) \in (0,\pi)$ such that
$\rho_+(\lambda) = {e}^{i\omega}$; see \cite[Lemma~2.4]{DSS}. This
allows for a very explicit analysis of the group
$\tilde{G}(\lambda)$ in terms of $a(\lambda)$, $b(\lambda)$ and
$\omega$. From $b(\lambda)\not= 0$ one can conclude
non-compactness, and $D(\lambda) \not= 0$ means $\omega \not=
\pi/2$, implying (\ref{3dir}). We refer to \cite{DSS} for details.
\end{proof}

\section{Proof of Localization via Multiscale Analysis}

\vspace{.2cm}

The proofs of Theorems~\ref{exploc} and \ref{dynloc} involve the
technical machinery of multiscale analysis, including the
verification of the necessary ingredients which are known to make
this whole apparatus work. Still, given the results which were
established in the previous sections, we can conclude with a brief
sketch of the remaining arguments. This is due to the fact that,
from here on, the strategy which was developed by Carmona et al.\
\cite{CKM} for the special case of the Anderson model can also be
followed for the more general random word models studied here. The
main point is that one has to stay away from the finite set of
exceptional energies.

First of all, for energy intervals in which positivity of $\gamma$
can be established through Furstenberg's theorem, that is, where
the group $G(\lambda)$ is non-compact and strongly irreducible, an
analysis of the $\lambda$-dependence can be performed which yields
H\"older continuity of $\gamma_0$, and thus $\gamma$. Details on
this, which other than smooth dependence of the transfer matrices
on $\lambda$ uses only general facts from the theory of products
of independent $SL(2,\R)$-matrices, can be found in \cite{CKM} and
\cite{Carmona/Lacroix}; see also \cite{DSS}, where it is shown
that this analysis also applies to continuum models.

Through the well known connection of $\gamma$ and the integrated
density of states (IDS) of ergodic discrete Schr\"odinger
operators provided by the Thouless formula, one then obtains
H\"older continuity of the IDS.

Combined with the positivity of $\gamma$, the latter is the basis
for proving a {\it Wegner estimate} as well as an {\it initial
length scale estimate}, both away from exceptional energies,
suitable for starting the multiscale analysis. The details of this
are straightforward extensions of the arguments provided in
\cite{CKM} for the Anderson model.

That multiscale analysis, given Wegner and initial length scale
estimates, proves exponential localization as claimed in
Theorem~\ref{exploc}, has been known since the 1980's and is
stated in \cite{CKM} together with all the fundamental references.
Improved versions, which show that strong dynamical localization
as in Theorem~\ref{dynloc} is obtained through multiscale
analysis, were recently provided in \cite{Damanik/Stollmann} and
\cite{Ger/Klein}. While these papers are written for continuum
operators, they both note that their results apply to lattice
models as well.

\setcounter{section}{1} \setcounter{equation}{0}
\renewcommand{\thesection}{\Alph{section}}
\renewcommand{\theequation}{\thesection.\arabic{equation}}

\section*{Appendix}

The aim of this appendix is to prove Proposition \ref{ergprop}. We
will start by providing a semi-algebra which generates the
$\p$-measurable sets and justifying that $\p$ is uniquely
determined by \eqref{defp}. We will then check that $T$, given by
\eqref{deft}, is measure-preserving and show the asserted
ergodicity properties by working on the semi-algebra; see Theorems
1.1 and 1.17 of \cite{Walters}. The semi-algebra we construct
consists of suitable cylinder sets. With these explicit sets we
are able to demonstrate that the main idea of the classical proof
of ergodicity of shifts, namely that two given cylinder sets
become independent if their non-trivial components are shifted
into disjoint regions, still holds. Due to the varying word
length, the details of this argument become more cumbersome, which
is our reason for including them here.

Let $\B_j$ denote the Borel sets in $\W_j$, $j=1, \ldots, m$, and
$\B$ the Borel sets in $\W$ (i.e., the $\nu$-measurable sets). The
set of all measurable cylinders of the form
\begin{equation} \label{cyl1}
\prod_{i= - \infty}^{-(n+1)} \W \times \prod_{i= -n}^{-1}A_i \times \widehat{A}_0
\times \prod_{i=1}^{n}A_i \times \prod_{i= n+1}^{\infty} \W,
\end {equation}
where $n \in \N$, $A_0 \in \B_j$, and $A_j \in \B$, for $1 \le |i| \le n$, is a
semi-algebra in $\{ \omega \in \Omega_0: | \omega_0 |=j \}$; denote it by
$\s_j$. Here the $\widehat{ \mbox{ }}$ symbol denotes the zero position in $\W^{\Z}$.
It follows that
\begin{equation} \label{semalg}
\s := \bigcup_{j=1}^m \left\{ \s_j \times \{ 1 \}, \ldots, \s_j \times \{ j \} \right\}
\end {equation}
is a semi-algebra in $\Omega = \bigcup_{j=1}^m \Omega_j$, where
$\Omega_j = \{ \omega \in \Omega_0: |\omega_0| = j \} \times \{1,
\ldots, j \}$.

We define a function $\p: \s \rightarrow \R^+$ by $\p(A \times \{
k \})= \p_0(A)/\langle L \rangle$ whenever $A \in \s_j$, and $1
\le k \le j$. $\p$ is countably additive since $\p_0$ is countably
additive and
\begin{equation} \label{pprob}
\sum_{j=1}^m \sum_{k=1}^j \p \left( \left( \prod_{i= - \infty}^{-1} \W \times \widehat{\W}_j
\times \prod_{i=1}^{\infty} \W \right) \times \{ k\} \right) =1.
\end {equation}
Thus $\p$ can be uniquely extended to a probability measure $\p$
on the $\sigma$-algebra $\Cal{F}$ in $\Omega$ generated by $\s$.
By construction, we have that $\p(A \times \{ k \}) =
\p_0(A)/\langle L \rangle$ for every $\p_0$-measurable $A \subset
\{ \omega \in \Omega_0: |\omega_0| =j \}$ and $1 \le k \le j$.
Define $T: \Omega \rightarrow \Omega$ by (\ref{deft}).
\begin{Lemma} \label{mpt}
$T$ is a measure-preserving bijection.
\end{Lemma}
\begin{proof}
$T$ is a bijection with
\begin{equation} \label{tinv}
T^{-1}(\omega,k) = \left\{ \begin{array}{ll} (\omega, k-1) & \mbox{if } k>1 \\
(T_0^{-1} \omega, |\omega_{-1}|) & \mbox{if } k=1. \end{array} \right.
\end {equation}
To prove that $T$ is measure-preserving it suffices to show that
$\p(T^{-1}M) = \p(M)$ for all $M \in \s$. Let $M=A \times \{ k
\}$, $A \in \s_j$, $1 \le k \le j$. Thus $\p(M) = \p_0(A)/\langle
L \rangle$.

If $k>1$, then $T^{-1}M = A \times \{ k-1 \}$ and $\p(T^{-1}M)= \p(M)$.

If $k=1$, represent $A$ in the form (\ref{cyl1}) and decompose
$$
A_{-1} = \bigcup_{j=1}^m C_j, \ \ C_j \in \B_j.
$$
Then, $T^{-1}(A \times \{ 1 \}) = \bigcup_{j=1}^m ( \Lambda_j \times \{ j \} )$, where
$$
\Lambda_j = \prod_{i= - \infty}^{-n} \W \times \prod_{i=
-n+1}^{-1}A_{i-1} \times \widehat{C}_j \times
\prod_{i=1}^{n+1}A_{i-1} \times \prod_{i= n+2}^{\infty} \W.
$$
Since $\p_0$ is additive and $T_0$ is measure-preserving, we get
$$
\p(T^{-1}(A \times \{ 1 \})) = \sum_{j=1}^m \p( \Lambda_j \times
\{ j\} ) = \sum_{j=1}^m \frac{ \p_0( \Lambda_j)}{\langle L
\rangle} = \frac{ \p_0(A)}{\langle L \rangle} = \p(A \times \{ 1
\}),
$$
concluding the proof.
\end{proof}

\begin{proof}[Proof of Proposition \ref{ergprop}.] (a) We assume that $J$ is
relatively prime. In order to show that $T$ is strongly mixing, we
need to prove that
\begin{equation} \label{stmix}
\lim_{ \ell \rightarrow \infty} \p(T^{-\ell}(A, k_A) \cap (B,k_B))
= \p(A,k_A) \cdot \p(B, k_B) = \frac{\p_0(A) \cdot
\p_0(B)}{\langle L \rangle^2}.
\end {equation}
Here $(A, k_A) \subset \Omega$ and $(B, k_B) \subset \Omega$ are arbitrary sets of the form
\begin{equation} \label{cyl2}
\begin{array}{c}
A= \cdots \times \W \times C_{-n} \times \cdots \times \widehat{C}_0 \times \cdots
\times C_n \times \W \times \cdots , \\
B= \cdots \times \W \times D_{-n} \times \cdots \times
\widehat{D}_0 \times \cdots \times D_n \times \W \times \cdots ,
\end{array}
\end {equation}
where $n \in \N$, $C_j \in \B_{a_j}$, $D_j \in \B_{b_j}$, $j=-n,
\ldots, n$, $1 \le k_A \le a_0$, and $1 \le k_B \le b_0$. The sets
$(A, k_A)$ with $A$ as in (\ref{cyl2}) are also a generating
semi-algebra $\tilde{\s}$ for $\Cal{F}$ (take finite disjoint
unions to get $\s$). Thus strong mixing follows from \eqref{stmix}
and \cite[Theorem~1.17~(iii)]{Walters}.

In order to calculate the left-hand side of \eqref{stmix}, we will
have to write $T^{-\ell}(A,k_A)$ as a disjoint union of sets
$(C,k)$ with $C$ of type \eqref{cyl1}. This is simple for $\ell =
\ell_0 +a_{-1}+ \cdots + a_{-n}-1$, where
\begin{equation} \label{shift}
T^{-\ell_0}(A,k_A) = (A_0,1),
\end {equation}
and
\begin{equation} \label{newcyl}
A_0= \cdots \times \W \times \widehat{C}_{-n} \times \cdots \times
C_n \times \W \times \cdots .
\end {equation}
For $\ell>\ell_0$, we have $T^{-\ell}(A,k_A) =
T^{-(\ell-\ell_0)}(A_0,1)$. Splitting sufficiently many of the
$\W$-factors in \eqref{newcyl} into their disjoint components
$\W_j$, we see that $T^{-(\ell-\ell_0)}(A_0,1)$ is a disjoint
union of sets of the form $( A(j_1, \ldots, j_r),k)$ where
\begin{equation} \label{Ajs}
A(j_1, \ldots, j_r) := \cdots \times \W \times \widehat{\W}_{j_1}
\times \cdots \times \W_{j_r} \times C_{-n} \times \cdots \times
C_n \times \W \times \cdots \, .
\end {equation}
We only need to determine the $( A(j_1, \ldots, j_r),k)$ in
$T^{-(\ell-\ell_0)}(A_0,1)$ with $k=k_B$, since otherwise the
intersection with $(B,k_B)$ is empty. Comparing \eqref{newcyl} and
\eqref{Ajs} and counting the length of the words shows that $(
A(j_1, \ldots, j_r),k_B) \subset T^{-(\ell-\ell_0)}(A_0,1)$ if and
only if $r$ and $j_1, \ldots, j_r$ are such that
\begin{equation} \label{conds}
k_B \le j_1 \le m, \ \ 1 \le j_2, \ldots, j_r \le m, \ \ j_1+
\ldots + j_r = k_B + (\ell-\ell_0)-1.
\end {equation}
This shows that
\begin{equation} \label{arbtran}
T^{-\ell}(A, k_A) \cap (B,k_B) = \bigcup (A(j_1, \ldots, j_r) \cap
B, k_B),
\end {equation}
where the disjoint union is taken over all $r, j_1, \ldots, j_r$
as in \eqref{conds}. If $\ell$ is sufficiently large, then we have
for all the sets on the right-hand side of \eqref{arbtran}, that
\begin{align*}
A(j_1, & \ldots, j_r) \cap B = \cdots \times \W \times D_{-n}
\times \cdots \times D_{-1} \times (\widehat{D_0 \cap \W_{j_1}})
\times \cdots
\\ & \cdots \times (D_n \cap \W_{j_{n+1}}) \times \W_{j_{n+2}}
\times \cdots \times \W_{j_r} \times C_{-n} \times \cdots \times
C_n \times \W \times \cdots.
\end{align*}
Hence,
$$
\p \left( T^{-\ell}(A, k_A) \cap (B,k_B)  \right) =
\frac{1}{\langle L \rangle^2} \p_0 \left( \bigcup (A(j_1, \ldots,
j_r) \cap B) \right) =
$$
$$
= \frac{1}{\langle L \rangle^2} \sum \p_0 \left( B \cap ( \cdots
\times \W \times \widehat{\W}_{j_1} \times \cdots \times \W_{j_r}
\times \W \times \cdots ) \right) \cdot \p_0(A).
$$
Taking into account further that
$$
B \cap ( \cdots \times \W \times \widehat{\W}_{j_1} \times \cdots
\times \W_{j_r} \times \W \times \ldots ) \neq \emptyset
$$
only if $(b_0, \ldots, b_n)=(j_1, \ldots, j_{n+1})$, in which case
$$
\p_0 \left( B \cap ( \cdots \times \W \times \widehat{\W}_{j_1}
\times \cdots \times \W_{j_{n+1}} \times \W \times \cdots )
\right) = \p_0(B),
$$
we conclude
\begin{equation} \label{almost}
\p \left( T^{-\ell}(A, k_A) \cap (B,k_B)  \right) =
\frac{1}{\langle L \rangle}\sum \nu_{j_{n+2}} \cdot \ldots \cdot
\nu_{j_r} \cdot \p_0(A)\p_0(B),
\end {equation}
where the sum runs over $1 \le j_{n+2}, \ldots, j_r \le m,
j_{n+2}+ \ldots + j_r = \ell -\ell_{A,B}$. Here we have set
$\ell_{A,B}:= \ell_0 + b_0 + \ldots +b_n +1-k_B$ and $\nu_j:=
\nu(\W_j)$. Thus, after an index shift, \eqref{stmix} becomes
equivalent to
\begin{equation} \label{conj}
\lim_{\ell \rightarrow \infty} \sum_{\footnotesize
\begin{array}{c} 1 \le j_1, \ldots, j_s \le m, \\ j_1+ \ldots +
j_s = \ell
\end{array}} \nu_{j_1} \cdot \ldots \cdot \nu_{j_s} =
\frac{1}{\langle L \rangle} = \frac{1}{\sum_{j=1}^m j \cdot
\nu_j}.
\end {equation}
Since $\sum_{j=1}^m \nu_j =1$ and $J$ is relatively prime, this is
exactly the result proven in Lemma \ref{rpconj} (a) below. This
finishes the proof of part (a).\\[1mm]
(b) If $J$ is not relatively prime, then by Lemma \ref{rpconj} (b)
below, the left-hand side of \eqref{conj} converges to $\langle L
\rangle^{-1}$ in Ces\`aro mean. Our previous arguments then yield
that
$$
\lim_{d \rightarrow \infty} \frac{1}{d} \sum_{\ell=1}^d \p \left(
T^{-\ell}(A, k_A) \cap (B,k_B)  \right) = \p (A,k_A) \cdot \p
(B,k_B),
$$
for all $(A,k_A)$ and $(B,k_B)$ in $\tilde{\s}$. By \cite[Theorem
1.17 (i)]{Walters} this implies ergodicity of $T$.

To complete the proof of part (b), it remains to check that in the
latter case, $T$ is not weakly mixing. To this end, note that by
the non-triviality condition (NC), there exist $j \in \{1, \ldots,
m \}$ and $C \in \B_j$ such that $0< \nu_j(C)<1$. Choose $A :=
\cdots \times \W \times \widehat{C} \times \W \cdots$, that is,
$A\times \{ 1\} \in \tilde{\s}$. If $\ell$ is not a multiple of
the greatest common divisor $D>1$ of $J$, then by considerations
as above one sees that
$$
\p \left( T^{-\ell}(A, 1) \cap (A,1)  \right) =0.
$$
Since $\p (A, 1) = \nu_j(C)/\langle L \rangle$, this implies that
$$
\liminf_{d \rightarrow \infty} \frac{1}{d} \sum_{\ell=1}^{d}
\left| \p \left( T^{-\ell}(A, 1) \cap (A,1) \right) -\p(A,1)^2
\right| \geq \frac{1}{2} \left( \frac{\nu_j(C)}{\langle L \rangle}
\right)^2 > 0.
$$
Thus $T$ is not weakly mixing; see \cite[Theorem 1.17
(ii)]{Walters}.
\end{proof}

In the above we have used a combinatorial lemma, which we state
and prove below. Let $m \in \mathbb{N}$ be fixed. For any $\ell
\in \mathbb{N}$, let the set of unordered partitions, or
compositions, of $\ell$ by natural numbers less than or equal to
$m$ be denoted by
\[ P(\ell,m) := \left\{ (j_1, \ldots, j_s) \in \{1,\ldots,m\}^s:
\: s\in \mathbb{N} \mbox{ and } \sum_{r=1}^s j_r = \ell \right\}, \]
and set ${\bf j_s}:=(j_1, \ldots, j_s) \in P(\ell,m)$.

\begin{Lemma} \label{rpconj}
For $j=1,\ldots,m$, suppose $0 \le \nu_j \le 1$ are given with
$\sum_{j=1}^m \nu_j =1$.\\
{\rm (a)} If $J := \{ j : \nu_j > 0 \}$ is relatively prime, then
\begin{equation} \label{lim1}
\lim_{\ell\to\infty} \sum_{ {\bf j_s} \in P(\ell,m) } \nu_{j_1}
\cdot \nu_{j_2} \cdot \ldots \cdot \nu_{j_s} =
\frac{1}{\sum_{j=1}^m j \cdot \nu_j}.
\end{equation}
{\rm (b)} If $J$ is not relatively prime, then
\begin{equation} \label{lim2}
\lim_{d\to\infty} \frac{1}{d} \sum_{\ell=1}^d \sum_{ {\bf j_s} \in
P(\ell,m) } \nu_{j_1} \cdot \nu_{j_2} \cdot \ldots \cdot \nu_{j_s}
= \frac{1}{\sum_{j=1}^m j \cdot \nu_j}.
\end{equation}
\end{Lemma}

\begin{proof} For each $\ell \in \mathbb{N}$, define
\begin{equation} \label{an}
A_{\ell} := \sum_{ {\bf j_s} \in P(\ell,m) } \nu_{j_1} \cdot
\nu_{j_2} \cdot \ldots \cdot \nu_{j_s},
\end{equation}
and for any $z \in \mathbb{C}$ with $|z|<1$, set
\begin{equation} \label{phit}
\tilde{\phi}(z) := \sum_{\ell=1}^{\infty} A_{\ell} z^\ell.
\end{equation}
Clearly,
\begin{align*}
\tilde{\phi}(z) & = \sum_{\ell=1}^{\infty} \sum_{ {\bf j_s} \in
P(\ell,m) } \nu_{j_1} \cdot \nu_{j_2} \cdot \ldots \cdot \nu_{j_s}
z^\ell \\
& = \sum_{s=1}^{\infty} \sum_{j_1=1}^m \ldots \sum_{j_s=1}^m
\nu_{j_1} \cdot \nu_{j_2} \cdot \ldots \cdot \nu_{j_s} z^{
\sum_{r=1}^s j_r } \\
& = \sum_{s=1}^{\infty} \left( \sum_{j=1}^m \nu_j z^j \right)^s\\
& = \frac{\sum_{j=1}^m \nu_j z^j}{1-\sum_{j=1}^m \nu_j z^j} \, .
\end{align*}
Take $A_0:=0$ and consider
$$
\phi(z):= (1-z) \tilde{\phi}(z).
$$
It is clear that both $\tilde{\phi}$ and $\phi$ are analytic in $|z|<1$.

Observe that the set $J:= \{ j: \nu_j>0 \}$ is relatively prime if and only if the
equation $\sum_{j=1}^m \nu_j z^j =1$ has exactly one solution ($z=1$) on the
unit circle.

We may conclude that if $J$ is relatively prime, then $\phi$ is
analytic in a neighborhood of $\{z \in \mathbb{C} : |z| \le 1 \}$,
and
$$
\lim_{\ell \rightarrow \infty} A_{\ell} = \phi(1) = \frac{1}{\sum_{j=1}^m j \cdot \nu_j}.
$$
This completes the proof of (a).

If the set $J$ is not relatively prime, then let $D$ be the
greatest common divisor of $J$, that is, suppose each $j \in J$
can be written as $j=D k_j$, and let $K := \{ k_j : j = Dk_j \in J
\}$.

Rescale the $\nu_j$ as follows: for each $1 \le j \le m$, define
$$
\tilde{ \nu}_j := \left\{ \begin{array}{ll} 0 & \mbox{if } j \notin K \\
\nu_{D j} & \mbox{if } j \in K. \end{array} \right.
$$
Note that the set $\tilde{J} := \{ j : \tilde{\nu}_j>0 \}$ is relatively prime by
construction. Taking
$$
\tilde{A}_{\ell} := \sum_{ {\bf j_s} \in P(\ell,m) }
\tilde{\nu}_{j_1} \cdot \tilde{\nu}_{j_2} \cdot \ldots \cdot
\tilde{\nu}_{j_s},
$$
one sees that
$$
A_{\ell} = \left\{ \begin{array}{ll} 0 & \mbox{if } \frac{\ell}{D} \notin \mathbb{N} \\
\tilde{A}_j & \mbox{if } \frac{\ell}{D} = j \in \mathbb{N}.
\end{array} \right.
$$
Thus
$$
\lim_{d \rightarrow \infty } \frac{1}{d} \sum_{\ell=1}^d A_{\ell} =
\lim_{d \rightarrow \infty } \frac{1}{d} \sum_{j \in \mathbb{N} : Dj \le d} A_{Dj}.
$$
Letting $m = d/D$, we have that
$$
\lim_{d \rightarrow \infty } \frac{1}{d} \sum_{\ell=1}^d A_{\ell} =
\frac{1}{D} \lim_{m \rightarrow \infty } \frac{1}{m} \sum_{j = 1}^m \tilde{A}_{j}
= \frac{1}{\sum_{j=1}^m j \cdot \nu_j},
$$
which completes our proof.
\end{proof}


\baselineskip=12pt
\begin{thebibliography}{99}

\bibitem{Bi/Ger} S.\ de Bi\`{e}vre and F.\ Germinet, Dynamical
localization for the random dimer Schr\"{o}dinger operator,
\textit{J.\ Stat.\ Phys.} {\bf 98} (2000), 1135--1148

\bibitem{Bougerol/Lacroix} P.\ Bougerol and J.\ Lacroix,
\textit{Products of Random Matrices with Applications to
Schr\"odinger Operators}, Birkh\"auser, Boston--Stuttgart (1985)

\bibitem{CKM} R.\ Carmona, A.\ Klein, and F.\ Martinelli, Anderson localization for
Bernoulli and other singular potentials, \textit{Commun.\ Math.\
Phys.} {\bf 108} (1987), 41--66

\bibitem{Carmona/Lacroix} R. Carmona and J. Lacroix,
\textit{Spectral Theory of Random Schr\"odinger Operators},
Birkh\"auser, Basel--Berlin (1990)

\bibitem{d2} D.\ Damanik, Gordon-type arguments in the spectral
theory of one-dimensional quasicrystals, in \textit{Directions in
Mathematical Quasicrystals}, Eds.~M.~Baake, R.~V.~Moody, CRM
Monograph Series {\bf 13}, AMS, Providence, RI (2000), 277--305

\bibitem{DSS} D.\ Damanik, R.\ Sims, and G.\ Stolz, Localization
for one-dimensional, continuum, Bernoulli-Anderson models,
\textit{Duke Math.\ J.} {\bf 114} (2002), 59--100

\bibitem{ds} D.\ Damanik and B.\ Solomyak, Some high-complexity
Hamiltonians with purely singular continuous spectrum, \textit{Ann.\
Henri Poincar\'{e}} {\bf 3} (2002), 99--105

\bibitem{Damanik/Stollmann} D.\ Damanik and P.\ Stollmann, Multi-scale
analysis implies strong dynamical localization, \textit{Geom.\ Funct.\
Anal.} {\bf 11} (2001), 11--29

\bibitem{VDK} H.\ von Dreifus and A.\ Klein, A new proof of localization in the
Anderson tight binding model, \textit{Commun.\ Math.\ Phys.} {\bf
124} (1989), 285--299

\bibitem{Ger/Klein} F.\ Germinet and A.\ Klein, Bootstrap multiscale analysis and
localization in random media, \textit{Commun.\ Math.\ Phys.} {\bf 222} (2001), 415--448

\bibitem{JSS} S.\ Jitomirskaya, H.\ Schulz-Baldes, and G.\ Stolz, Delocalization in
random polymer models, preprint (2002), available at mp-arc 02-267

\bibitem{Kato} T. Kato, \textit{Perturbation Theory for Linear Operators},
Springer-Verlag, Berlin--Heidelberg (1966)

\bibitem{Kirsch} W.\ Kirsch, On a class of random Schr\"odinger operators, \textit{Adv.\
in Appl.\ Math.} {\bf 6} (1985), 177--187

\bibitem{Knopp} K. Knopp, \textit{Theory of Functions}, Vol.
II, Dover, New York (1947)

\bibitem{Petersen} K.\ Petersen, \textit{Ergodic theory}, Cambridge University Press,
Cambridge (1983)

\bibitem{SVW} C.\ Shubin, R.\ Vakilian, and T.\
Wolff, Some harmonic analysis questions suggested by
Anderson-Bernoulli models, \textit{Geom.\ Funct.\ Anal.} {\bf
8} (1998), 932--964

\bibitem{t} G.\ Teschl, \textit{Jacobi Operators and Completely
Integrable Nonlinear Lattices}, Mathematical Surveys and
Monographs {\bf 72}, AMS, Providence, RI (2000)

\bibitem{Walters} P.\ Walters, \textit{An Introduction to Ergodic
Theory}, Springer-Verlag, New York--Berlin (1982)

\end{thebibliography}
\end{document}